\documentstyle[12pt]{article}
\def\lsim{\mathrel{\rlap {\raise.5ex\hbox{$ < $}}
{\lower.5ex\hbox{$\sim$}}}}
\def\gsim{\mathrel{\rlap {\raise.5ex\hbox{$ > $}}
{\lower.5ex\hbox{$\sim$}}}}

\newcommand{\pr}{\paragraph{}}
\newcommand{\be}{\begin{equation}}
\newcommand{\ee}{\end{equation}}
\newcommand{\bea}{\begin{eqnarray}}
\newcommand{\nn}{\nonumber}
\newcommand{\eea}{\end{eqnarray}}
\newcommand{\nd}[1]{/\hspace{-0.6em} #1}
\newcommand{\nk}{\noindent}
\def\gappeq{\mathrel{\rlap {\raise.5ex\hbox{$>$}}
{\lower.5ex\hbox{$\sim$}}}}

\def\lappeq{\mathrel{\rlap{\raise.5ex\hbox{$<$}}
{\lower.5ex\hbox{$\sim$}}}}

\begin{document}

\begin{titlepage}
\begin{flushright}
OUTP-95-32P \\
\end{flushright}

\begin{centering}
\vspace{.1in}
{\large {\bf Deviations from Fermi-Liquid behaviour in
(2+1)-dimensional Quantum Electrodynamics and the
normal phase of
high-$T_c$ Superconductors }} \\
\vspace{.2in}
{\bf I.J.R. Aitchison } and
{\bf N.E. Mavromatos}$^{*}$ \\
\vspace{.1in}
\vspace{.03in}
Department of Physics
(Theoretical Physics), University of Oxford, 1 Keble Road,
Oxford OX1 3NP, U.K.  \\

\vspace{.1in}
{\bf Abstract} \\
\vspace{.05in}
\end{centering}
{We argue that the gauge-fermion interaction
in multiflavour quantum electrodynamics in $(2 + 1)$-dimensions
is responsible for non-fermi liquid behaviour in the infrared,
in the sense of leading to the existence of a non-trivial
(quasi) fixed point that lies
between the trivial fixed point (at infinite momenta)
and the region where dynamical symmetry breaking
and mass generation occurs. This quasi-fixed point
structure implies slowly varying, rather than fixed, couplings
in the intermediate regime of momenta, a situation which
resembles that of (four-dimensional) `walking technicolour' models of
particle physics.
The
inclusion of wave-function renormalization
yields marginal $O(1/N)$-corrections
to the `bulk' non-fermi liquid behaviour
caused by the gauge interaction in the limit of
infinite flavour number. Such corrections
lead to the appearance of modified critical
exponents.
In particular, at low temperatures there appear to be
logarithmic scaling violations of the linear resistivity
of the system of order $O(1/N)$.
Connection with
the anomalous normal-state properties of certain
condensed matter systems relevant for
high-temperature
superconductivity is briefly discussed. The
relevance of the large (flavour) $N$ expansion
to the fermi-liquid problem is emphasized.
As a partial result of our analysis, we
point out the
absence of Charge-Density-Wave Instabilities
from the effective low-energy theory,
as a consequence of gauge invariance.}
\vspace{0.2in}
%
\pr

\vspace{0.01in}
\begin{flushleft}
$^{*}$P.P.A.R.C. Advanced Fellow \\
September 1995 \\
\end{flushleft}
\end{titlepage}
\newpage
\section{Introduction}
\pr
One of the most striking phenomena associated
with the novel high-temperature superconductors
is their {\it abnormal} normal-state properties.
In particular, these substances are known to exhibit
deviations from the known Fermi-liquid behaviour,
which are remarkably stable with respect to
variations in the relevant parameters~\cite{nonfermi}.
Recently, Shankar~\cite{rgshankar} and Polchinski~\cite{polch}
have presented an intuitively appealing idea
of using the Renormalization-Group (RG) approach, so powerful
in particle and statistical physics, to systems of interacting
electrons with a Fermi surface in order to understand, at least
qualitatively, how deviations from Fermi liquid
behaviour can appear {\it naturally} (as opposed to being
fine-tuned).
{}From this point of view Landau's fermi liquid
is nothing else but a system of free electrons, which has
no relevant perturbations, in the RG sense,
that can drive it away from its trivial infrared
fixed point. In general, however,
as we integrate out certain modes of our original theory, some
interactions may become relevant in the RG sense,
i.e. their effective coupling may grow as one lowers
the momentum scale. Then, two interesting possibilities
arise~\cite{polch}. (i) Fermion bound states are formed, symmetries
are spontaneously broken, and the low-energy spectrum bears little resemblance
to that of the original theory. In such a case one has to re-write
the effective theory in terms of the new degrees of freedom :
for instance, in the superconducting case this is the
Landau-Ginzburg effective action expressed in terms of the fermion
condensate. (ii) Alternatively, the growth of the coupling
is cut off by quantum
effects
at a certain low energy scale, and in this way a
{\it non-trivial} fixed point structure emerges.
The low energy fluctuations still correspond
to fields of the original theory despite their
non-trivial interactions. This case leads
to observable deviations from
the Fermi-liquid behaviour.
\pr
In the case of the high-$T_c$ materials, the physically
interesting question is whether one model theory
can be found with a structure rich enough
to describe {\it both} the non-fermi liquid behaviour
of the normal phase {\it and} the transition to
(and phenomenology of) the superconducting phase. In this
article we shall put forward a candidate
model which, as we shall argue, seems to us to fulfill this r\^ole.
\pr
It is known that possibility (i) above can be caused by relevant interactions
of superconducting (BCS) or charge-density-wave (CDW) type,
both of which are accompanied by the formation of fermion condensates.
Possibility (ii) has only rather recently begun to be seriously
explored~\cite{rgshankar,polch,nayak}. It has been known for a long time
that the electromagnetic interaction of the vector potential
can cause deviation from fermi-liquid behaviour~\cite{vanalphen},
but its effects are suppressed by terms of $O[(v_F/c)^2]$,
with $v_F$ the Fermi velocity and $c$ the light velocity.
Its effects occur only
at much lower energies than those relevant to the high-$T_c$
materials. Nevertheless, the electromagnetic example
is suggestive enough,
perhaps, to motivate a search for other (non-electromagnetic)
gauge interactions in which the effective
signal velocity would be of order $v_F$, and which
might be responsible for a non-trivial fixed point behaviour.
It was precisely this sort of (``statistical'')
gauge-fermion interaction that was studied
(in different forms) in \cite{polch} and \cite{nayak},
and which led to non-trivial fixed point structure
in the infrared.
\pr
Returning
now to possibility (i), we recall that
it has been shown~\cite{doreym}
that a variant of $QED$ in (2 + 1)-dimensions ($QED_3$) leads
to superconductivity, characterized - as appropriate to two space dimensions -
by the absence of a local order parameter (Kosterlitz-Thouless
mode). Thus the exciting possibility arises that a single fermion-gauge
theory could describe both non-fermi-liquid
behaviour in the normal phase and the transition to the
superconducting phase.
\pr
Formulated
in terms of $N$ species of electromagnetically charged fermions,
the model of \cite{doreym} (to which we shall return in section 4)
consists of a $CP^{N-1}$ $\sigma$-model coupled to the fermions
via the gauge field of the $\sigma$-model representing
magnetic spin-spin interactions.
The main purpose of the
present article
is to present an (approximate) renormalization group analysis
of a simplified version
of this model, namely $QED_3$ itself, which indicates
that $QED_3$ exhibits two quite
different behaviours depending on the momentum scale.
At very low momenta $QED_3$
enters a
regime of dynamical mas generation (d.m.g.), which in the full theory
leads to superconductivity;
but at ``intermediate'' momenta (see below) d.m.g. does not
occur and the dynamics
is controlled by a non-trivial
fixed point, leading to non-fermi liquid behaviour. Thus we have
the possibility - for the first time, to our knowledge -of one theory
encompassing both the normal and the superconducting phases of
the high-$T_c$ cuprates.
\pr
We postpone until section 4 a fuller account of the realistic
model we are advocating. Before that, in Sections 2 and 3,
we shall consider for clarity the simpler case of $QED_3$,
which as we shall see already exhibits the crucial dynamical features
(however, as we shall see in section 4, $QED_3$ describes only
a part of the realistic model believed to simulate
the physics of the high-$T_c$ cuprates).
{}From this we conclude that the essential dynamical ingredient
in our model is simply that it is a {\it $U(1)$ gauge theory
in two space dimensions }.
\pr
At this point the reader might worry that applying
renormalization group techniques to a super-renormalizable
theory like $QED_3$ is redundant, since the theory
has no ultraviolet divergencies.
However, this is a mistaken view.
In the modern approach to the RG and effective field theories, one
considers quite generally how a theory evolves
as one integrates out degrees of freedom above
a certain momentum scale, moving progressively down in scale.
{}From this point of view an effective field
theory description is equally applicable to non-renormalizable,
renormalizable, and super-renormalizable theories.
However, there are some crucial
new features in the case of a super-renormalizable
theory (which, to our knowledge, have not
been identified hitherto). First, the $QED_3$ coupling $e$
introduces an intrinsic {\it intermediate}
scale $e^2$ which has the dimension of mass, this being directly
related to the super-renormalizability of the theory. The physical
effect of this will be the existence of an intrinsic
mass scale and we can expect different physics in different
regimes of momenta relative to this mass scale ($p >> e^2$,
$p \simeq e^2$, $p << e^2$).
\pr
The second distinctive feature of our RG analysis of $QED_3$,
concerns the way in which we introduce a running coupling. Conventionally,
such running couplings are dimensionless - so, once again the
dimensionfulness of $e^2$ presents a new feature. The way in
which an effective dimensionless running coupling can be introduced into
$QED_3$ has been shown
by Kondo and Nakatani (KN)~\cite{kondo}, building on work by
Higashijima~\cite{higashijima} for $QCD_4$.
The crucial step is to consider the effect of wavefunction
renormalization in the Schwinger-Dyson (SD) equations, as controlled
by a large-$N$ approximation. In this case, one considers
the theory at large $N$ with $\alpha =e^2 N$ held fixed,
and the dimensionless coupling that runs is essentially
$1/N$.
\pr
KN actually considered only the regime in which dynamical mass generation
(chiral symmetry breaking) occurs - and of course here the gauge coupling
is becoming strong and the use of a large-$N$ expansion
is unavoidable. What we shall do (in section 2) is to
identify the ``normal'' (no dynamical mass generation)
regime of the theory, and extend the RG-type analysis of KN to this
normal regime.  We shall argue that there exists a non-trivial
fixed point of the effective dimensionless coupling, which governs
the dynamics for a range of {\it intermediate}
momenta $p \simeq \alpha $, lying between
the trivial fixed point at $p >> \alpha $, and the region
$p << \alpha $ of dynamical mass generation.
Important to this analysis will be the introduction
(following KN) of an infrared cutoff $\epsilon $, which
serves to delineate the different momentum regimes.
\pr
The analysis of section 2 is performed at zero temperature, and in section 3
we shall try to connect this to finite-temperature
calculations, by interpreting
the temperature as an effective infrared cutoff. We present
an approximate computation, at finite temperature, of the
electrical resistivity
$\rho $ of the fermionic system.
We argue that it is the existence of the non-trivial RG fixed point
which is responsible for the fact
that the non-fermi liquid behaviour
($\rho $ approximately proportional to the
temperature $T$) is observed
over so large a temperature range. Wavefunction
renormalization effects, important at $O(1/N)$, lead to
calculable logarithmic deviations from the linear in $T$ behaviour.
\pr
Before proceeding further, it is useful
to compare and contrast our approach with two other
recent explorations of gauge theories in $(2 + 1)$ dimensions
in a similar context, by Polchinski~\cite{polch} and by
Nayak and Wilczek~\cite{nayak}. Both works deal with
fermions interacting with a statistical gauge field,
the latter representing magnetic spin-spin
interactions (as in our $CP^{N-1}$
sector, see section 4). In both,
the fermions represent {\it spin} quasi-particle
excitations (spinons), and they should therefore
not be identified with the carriers of ordinary
electric charge (holes or electrons). This is to be
sharply contrasted with our own model of section 4, in
which the spin-charge separation is done differently,
leading to the fermions in our model carrying both
statistical and ordinary charge.
\pr
The model of ref. \cite{nayak} consists of a gauge-fermion
interaction,
in the presence of
a modified four-fermion interaction
of a long-range $1/k^x$ form, with $k$ the momentum.
An important r\^ole is also played by a $P$-
and $T$- violating term,
in the form of a Chern-Simons interaction for the gauge
field. The latter is responsible for enslaving gauge field
fluctuations to density fluctuations.
In the case $x < 1$ this
results in a relevant gauge-fermion interaction.
Nayak and Wilczek~\cite{nayak} have shown, by
employing a systematic expansion in powers of $1 - x$,
the existence of a non-trivial infrared fixed point
responsible for deviations from Fermi liquid
behaviour.
The importance of the Chern-Simons interaction
lies in the fact that it allows, through the constraint
implied by integrating out the temporal component of the
statistical gauge field,
a rewriting of the non-local $1/k^x$-four-fermi interaction
as a Maxwell-like term for the gauge field
but with modified $1/k^x$ momentum behaviour. The ordinary
Maxwell term corresponds to $x=0$, whilst the
Coulomb interaction corresponds to $x=1$.
Up to its non-relativistic form, which is a consequence of
the non-relativistic character of the fermion-gauge system
with a fermi surface, this situation is qualitatively similar
to the dimensional reduction of the ordinary
Maxwell term from four to three space-time dimensions~\cite{doreym}.
Indeed in that case, a three dimensional Maxwell term for the
electromagnetic field $A_M$, $M=1,2,3$,
corresponding to the projection of a four-dimensional
theory
onto the spatial plane, results in a Coulomb-like
form for the gauge field kinetic term
\be
    \int d^3x F_{MN}(A)\frac{1}{\sqrt{\nabla ^2}}F^{ MN} (A)
\label{maxwell}
\ee
This result is due to the fact that in three space-time
dimensions the Green's functions for the dimensionally-reduced
Maxwell field
are modified appropriately to yield the above
`square-root-of-$\nabla ^2$' behaviour (\ref{maxwell}).
It is natural, therefore, to imagine that a behaviour
$(\sqrt{\nabla ^2})^{-1+\epsilon }$ may be attributed to
quasi-planar geometries, or to deviations from three space-time
dimensions as in dimensional regularization $D=3 + \epsilon $
with $\epsilon = 1$ corresponding to
the (Maxwell) four-dimensional
kinetic term.
\pr
{}From this analogy one can understand that
the parameter $1-x$ of ref. \cite{nayak}
plays a r\^ole similar to that of
the $\epsilon$ parameter
of Wilson or of dimensional regularization.
This is the advantage of the method of ref. \cite{nayak},
in the sense
of providing a controlled expansion in powers of $1 -x$,
which can lead to a non-trivial fixed point for the
gauge-fermion interaction at weak coupling.
\pr
The above work, makes explicit use of Parity (P) and Time-Reversal
(T) breaking effects of the ground state, which, however, is difficult
to reconcile with experiment at present.
To avoid this difficulty,
Polchinski~\cite{polch}
examined the possibility of a non-trivial
infrared fixed point in a P and T conserving situation
in which the only non-trivial interaction
in the effective lagrangian of spinons is that
with the statistical gauge field without any
Chern-Simons term.
This is formally the same
as the essential fermion-gauge sector of our own model,
but with the crucial physical difference -to repeat -
that our fermions will (in sections 3 and 4) carry electric charge,
whereas Polchinski's cannot. To have a controllable
expansion
Polchinski~\cite{polch} employed
a large $N$ analysis in the fermionic flavours
by extending the $SU(2)$ spin group to $SU(N)$, $N \rightarrow
\infty$. He presented a Schwinger-Dyson analysis
for the propagators of the fermion and gauge fields,
which he solved in a closed form to leading order
in the $1/N$ expansion by invoking a tree-level ansatz
for the gauge-fermion vertex at large $N$
at low energies. Renormalization, then, implies that
the gauge-fermion interaction is promoted from
irrelevant to {\it marginal}, thereby sowing the possibility
of a non-trivial fixed point of this model in the
infrared and, hence, its non-fermi-liquid behaviour.
Because the kinetic term for the gauge field
assumes the normal Maxwell form, the results of Polchinski
can probably
be classified as belonging to the $x=0$
universality class in the language of
Nayak and Wilczek~\cite{nayak}.
The criticism that one may make of Polchinski's approach
is the fact that he neglects renormalization effects on the
vertex, which can lead to a non-consistent expansion
in $1/N$. Such effects were crucial in the work of
Nayak and Wilczek in order to get a controllable
expansion in the fermion
self-energy calculation at (resummed) one loop.
\pr
The important observation in Polchinski's work, which will
be directly relevant for our purposes here, is that
kinematics implies that the most important
interactions among fermions
are those which pertain to fermionic excitations
whose momentum components tangent to the fermi surface are
parallel. This is the only way that the gauge field momentum
transfer can still be relatively large as compared to the
distance of the fermion momenta from the fermi surface, as required
by special kinematic conditions~\cite{polch}.
There are two cases where such conditions are met
in condensed matter physics. The first pertains to
nested fermi surfaces, at which the points with momenta
$k_0$ and $-k_0$ have parallel tangents. This is
the situation relevant to  BCS or CDW.
The other situation, which is the bulk of Polchinski's
work and will be of interest to us as well, is the case
where the fermions are close to a single point
on the fermi surface.
This means that the most important fermion interactions
are those which are local on the fermi surface, and hence
qualitatively  this situation can be extended to
relativistic (Dirac) fermions  as well, since the
dispersion relations become effectively linear~\cite{doreym}.
\pr
Another important point, which was recently pointed out
by Shankar~\cite{rgshankar} in connection with the
RG approach to interacting fermions,
is the use of an effective large-$N$ expansion
in cases where the effective momentum cut-off
$\Lambda $ is much
smaller than the size of the fermi surface   $k_F$,
$\Lambda /k_F \rightarrow 0$.
Such a situation is encountered in
a RG study of (deviations from) fermi liquid theories,
the Landau fermi-liquid theory being defined as a trivial
infrared fixed point in a RG sense.
To understand the connection of a large-$N$ expansion with
infrared behaviour of excitations
one should recall the work of ref. \cite{gallavoti}
where the RG approach to the theory of the Fermi surface
has been studied in a mathematically rigorous way.
The basic observation of ref. \cite{gallavoti} is that,
unlike
the case of relativistic field theories,
in systems with an extended fermi surface,
the fermionic excitation
fields exhibiting the correct scaling are not the original
excitations, $\psi _x$ ($x$ a configuration space variable),
but rather {\it quasiparticle} excitations
defined  as follows :
\be
    \psi _x= \int _{|{\bf \Omega} |=1} d{\bf \Omega} e^{ik_F {\bf \Omega}
. {\bf x}}
\psi _{x,\Omega} = \int _{|{\bf \Omega} |=1} d{\bf \Omega}
e^{i(k_F {\bf \Omega} -{\bf K}). {\bf x} } {\tilde \psi}_{{\bf K,\Omega}}
\label{relgal}
\ee
where for the shake of simplicity we assumed that the fermi surface
is spherical with radius $k_F$,
${\bf \Omega}$ is a set of angular variables defining
the orientation of the momentum vector of the excitation
at a point on the fermi surface, and
the tilde denotes ordinary Fourier transform in a momentum space ${\bf K}$.
These quasiparticle fields
have propagators with the correct scaling~\cite{gallavoti},
which allows ordinary RG techniques, familiar from relativistic
field theories, to be applied, such as the appearance of
renormalized coupling constants, scaling fields etc.
Indeed it is not hard to understand why this is so.
For this purpose it is sufficient to observe
that for large $k_F$
the exponent of the exponential
in (\ref{relgal}) is nothing other
than the {\it linearization},
${\bf k} \equiv {\bf K}- k_F {\bf \Omega }$,
about a point on the fermi surface, which makes
these quasiparticle excitations identifiable
with ordinary field variables of the low-energy limit
of these condensed matter systems. The latter is a well-defined
field theory~\cite{doreym}.
The crucial point in this interpretation is that
now the field variables will depend on `internal
degrees of freedom', ${\bf \Omega}$,
which denote angular orientation of the momentum vectors on the
fermi surface. In two spatial dimensions, which is the case of
interest, ${\bf \Omega}$ is just the polar angle $\theta$.
Following ref. \cite{rgshankar}
we discretize this angular space into small cells of extent
$f(\Lambda/k_F) << 1$, e.g. $f=\Lambda/k_F$:
\be
\int \frac{d^2k}{4\pi^2} \equiv
\int _{-\Lambda}^\Lambda \frac{dk}{2\pi}
\int_{f(\Lambda/k_F)}^{f(\Lambda/k_F)} k_F
\frac{d\theta}{2\pi}
\label{integration}
\ee
where $k$ denotes a linearizing momentum about a point
on the fermi surface.
Doing so, we observe~\cite{rgshankar}
that when looking at interaction
terms involving fermionic particle-antiparticle pairs,
${\overline \psi}\psi$,
the leading interactions are among those fermion-antifermion pairs
for which the creation and anihilation
operators lie within the same
angular cell. This is for purely kinematic reasons in the infrared regime
$\Lambda << k_F$, similar to those
mentioned previously~\cite{polch},
which implied that the most
important fermion interactions on the fermi surface must
be among excitations which have their
tangents to the fermi surface
parallel. It is, then, straightforward to see
that interaction terms involving either gauge excitations or
just fermions resemble those in large-$N$
relativistic field theories, given that
the only $\Lambda$ dependence appears through proportionality
factors $f(\Lambda/k_F) << 1$
in front of the interactions, in the infrared.
One, then, identifies $1/N$ with $f(\Lambda/k_F) << 1$,
and the only
difference from ordinary particle-physics large-$N$ expansions
is the dependence of this effective $N$ on the cut-off $\Lambda$:
that is to say, $1/N$ runs.
\pr
As we shall show  in the next section, however,
large $N$ expansions in three dimensional $QED$ can exhibit
such scale dependence.
Wave-function renormalization
leads to a renormalized
`running' $1/N$. Furthermore, the running is of a
novel nature.
Instead of finding a non-trivial
infrared fixed point, we shall demonstrate the
existence of an (intermediate) regime of momenta, where
the effective running of the gauge coupling, which is essentially
$1/N$ times a spontaneously appearing scale, is slowed down
considerably, so that one encounters a quasi-fixed-point
situation. As we shall argue, this quasi-fixed point
structure is sufficient to cause (marginal) deviations
from the fermi liquid picture.
In view of the above, this makes
such theories plausible candidates for a correct
qualitative description of deviations from
Landau fermi liquid theory. This has obvious relevance
to the normal phase properties of
(realistic) condensed matter systems~\cite{doreym},
advocated in section 4, which are believed to
simulate the physics of the high-$T_c$ cuprates.

\section{$QED_3$: Super-renormalizability,
`running' couplings and non-trivial (quasi-)fixed-point structure}

\subsection{Wave-function Renormalization and running flavour
number }
\pr
Three-dimensional quantum electrodynamics ($QED_3$)
has recently received a great deal of attention (\cite{app}-\cite{ait})
not only as a result of its potential application to the study of
planar high-temperature superconductivity \cite{doreym}, mentioned
in the introduction, but also because
of its use as a
protoype for studies of chiral symmetry breaking
in higher-dimensional (non-Abelian) gauge theories \cite{miran}.
\pr
However, despite the theory's apparent simplicity  the
situation is not at all clear at present.
A great deal of controversy has arisen
in connection with the r\^ole of the wave-function
renormalization. In the early papers \cite{app}
the wave-function renormalization $A(p)$ was argued to be
$1$
in Landau gauge
to leading order in $1/N$, where $N$ is the number of fermion
flavours, and thus was ignored.
More detailed studies, however, showed
\cite{pen} that the precise form, within the resummed $1/N$
graphs, of $A(p)$ is
\be
     A(p) = (\frac{p}{\alpha })^{\frac{8}{3N\pi ^2 }}
\label{one}
\ee
where $\alpha =e^2 N$ is the dimensionful coupling constant
of $QED_3$, which is kept fixed as $N \rightarrow \infty $.
It is clear from (\ref{one}) that, although at energies
$p \simeq \alpha $ the wave-function is of order one,
however at low momenta $p << \alpha $, relevant for
dynamical generation of mass, the wave-function renormalization
yields logarithmic scaling violations which could affect \cite{pen}
the
existence of a critical number of flavours $N_c$, below which,
as argued in ref. \cite{app}, dynamical mass generation occurs.
However, this result was not free of ambiguities
either, given that the inclusion of wave-function
renormalization necessitates the introduction of a non-trivial
vertex function. The exact expression for the latter is not
tractable, even to order $O(1/N)$, and one has to assume various
ansatzes \cite{pen} that can be questioned.
The situation became clearer after the work of ref. \cite{kondo},
who showed that the introduction of an infrared cut-off
affects the results severely, depending on the various
ansatzes used for the vertex function. In particular,
as the authors of ref. \cite{kondo} showed, there are extra
logarithmic scaling violations in the expression for $N_c$,
depending on the form of the vertex function assumed,
which render the limit where the infrared cut-off is removed,
not well-defined.
\pr
For our present purposes, however, we are not
so much interested in whether the inclusion
of wavefunction renormalization leads to a critical
$N_c$ or not, as in the more general point that - as
noted by Kondo and Nakatani (KN)~\cite{kondo}, following
Higashijima~\cite{higashijima} - the vacuum polarization
contribution to $A$ produces effectively
a running coupling, even in the case of the
super-renormalizable theory of $QED_3$. KN's analysis
was restricted to the regime of dynamical
mass generation, and our main purpose in this section
is to extend that to the ``normal'' regime where mass is not
dynamically generated. We emphasize now, however,
that if $A$ is set equal to unity at the outset, the power
of the running coupling concept to unify both regimes
is completely lost.
\pr
We therefore continue with a brief review of the analysis of
\cite{kondo}. Their vertex ansatz was assumed to be
\be
\Gamma _\mu (q,p) = \gamma _\mu A(p)^n \equiv \gamma _\mu G(p^2)
\label{three}
\ee
where $p$ denotes the momentum of the photon.
The Pennington and Webb \cite{pen} ansatz corresponds to
$n=1$, where chiral symmetry breaking occurs for
arbitrarily large $N$ \cite{pis}. It is this case that
was argued to be consistent with the
Ward identities that follow from gauge invariance \cite{pen}.
In this paper we shall concentrate on the
generalized
ansatz, with $n \ne 1$,
and in particular we shall discuss its finite temperature
behaviour.
We keep the exponent $n$
arbitrary \cite{kondo} and discuss qualitatively the
implications of the vertex ansatz for various ranges of the
parameter $n$. As we shall argue below this is crucial for
the low-energy renormalization-group structure of the model.
\pr
Using the ansatz (\ref{three}), Kondo and Nakatani~\cite{kondo}
proceeded to analyze the Schwinger-Dyson (SD) equations,
in the regime of dynamical mass generation, in terms of a running
coupling as follows. Their (approximate) SD
equation for $A(p)$ is (in Landau gauge)
\be
  A(p) = 1 - \frac{g_0}{3} \int _\epsilon ^\alpha dk
\frac{k A (k) G(k^2) }{k^2 A^2 (k) + B(k^2)}
\{(\frac{k}{p})^3 \theta (p - k) + \theta (k - p) \}
\label{sdir}
\ee
where $g_0 =8/\pi ^2 N$, $N$ is the number of fermion
flavours, and $\epsilon$ is an infrared cutoff.
In the low-momentum
region relevant for dynamical mass generation $p << \alpha $
and the first term in the right-hand-side of (\ref{sdir}),
cubic in $(\frac{k}{p})$,
may be ignored. Then, taking into account that $G(k^2)=A(k)^n$,
and using the bifurcation method in which one ignores the
gap function $B(k)$
in the denominators of the SD equations,
one obtains easily
\be
  A(t)=1-\frac{g_0}{3} \int _t^0 ds A^{n-1} (s)
\label{uvsd}
\ee
which has the solution
\be
 A(t) = (1 + \frac{2-n}{3}g_0 t)^{\frac{1}{2-n}}
\qquad ; \qquad
 t \equiv ln (p/\alpha )
\label{uvsol}
\ee
Substituting to the SD equation for the gap, one then obtains
a `running' coupling~\cite{kondo} in the low momentum region
\be
 g^L =\frac{g_0}{1 + \frac{2-n}{3}g_0 t }
\label{two}
\ee
which, we note, is actually independent of $\epsilon$.
The existence of the dimensionless parameter $g^L$
in $QED_3$ may be associated with the ratio
of the gauge coupling $e^2/\alpha  $, given that in the
large $N$ analysis the
natural dimensionful scale  $\alpha$
has been introduced. Thus, a renormalized
running $N^{-1}$ might be thought of expressing
`charge' scaling in this super-renormalizable theory.
In particular (\ref{two})
implies that the $\beta$ function
corresponding to  $g^L$ is of `marginal' form
\be
    \beta ^L \equiv -\frac{d g^L}{dt} = \frac{2-n}{3}(g^{L})^2
\qquad .
\label{betaf}
\ee
Thus, depending of the sign of $2 -n $ one might have
{\it marginally}
relevant or irrelevant couplings $g^L \propto e^2/\alpha$.
The first derivative of the
$\beta$ function with respect to the coupling $g^L$
is
\be
    \frac{d}{d g^L} (\beta ^L) =2\frac{2-n}{3}g^L
\label{derbet}
\ee
and since $g^L > 0$ by construction, its sign depends
on the sign of $n-2$.
For $n < 2$  (the marginally relevant case)
the gauge interaction decreases rapidly as one
moves away from low momenta,
and the theory is ``asymptotically free''~\cite{kondo}.
If $n > 2$ (marginally irrelevant), on the other hand, then
$g^L (t)$ tends to zero in the low momentum region, whilst for
$n=2$ the coupling is exactly marginal and one recovers the
results of ref. \cite{app,atkinson} about the existence of a critical
flavour number.
Gauge invariance, in the sense of the Ward-Takahashi identity seems to
imply~\cite{pen,atkinson}
$n \le 2$ and this is the range we shall explore in this article.
\pr
Our problem now is to extend (\ref{two}) beyond
the region $p << \alpha $. Consider first the true ultraviolet
region $p \rightarrow \infty $. Assuming for the moment
that (\ref{two}) were correct for $p >> \alpha $, one finds
a zero of the $\beta$ function
at the point $t \rightarrow  \infty $, the trivial
fixed point $g^* = 0$,
which is
an ultraviolet fixed point.
However, (\ref{two}) or (\ref{betaf})
are not reliable for the range of momenta
$p >> \alpha $. Both formulas
have been derived in the regime of momenta relevant
to the dynamical mass generation,
$p << \alpha $.
\pr
This being so, do we have an alternative argument
for a trivial ultraviolet fixed point?
The answer is affirmative.
To this end we use the results of ref. \cite{kondoquench}
employing a quenched fermion approximation in large $N$
QED. The result of such an investigation is that
once fermion loops  are ignored,
and hence only tree-level graphs (ladder) are
taken into account, the wave-function renormalization
is rigorously proved to be trivial in the Landau gauge~:
\be
      A(p)^{quenched}  = 1
\label{quenched}
\ee
This result is a consequence of special mathematical
relations of resummed ladder graphs in Schwinger-Dyson
equations. Now in our case, one observes that
in the high-energy regime, $p \rightarrow \infty$,
the $\frac{1}{N}$-resummed gauge-boson
polarization tensor vanishes as
$\Pi (p \rightarrow \infty ) \simeq \alpha /8 p \rightarrow  0 $.
Thus, the situation is similar to the quenched approximation,
which implies the absence of any wave-function
renormalization (\ref{quenched}), and therefore
the vanishing (triviality) of the effective (`running')
coupling constant $g$ in the ultra-violet regime of momenta.
This is in qualitative agreement with the
naive estimate made above, based on the
formulas (\ref{two}), (\ref{betaf}).
\pr
The situation is, therefore, as follows.
The coupling grows from the trivial fixed-point
(ultraviolet regime), where there is no mass-generation
to stronger values as the momenta become lower.
According to the naive formula (\ref{betaf}),
this coupling grows indefinitely for low momenta
and the perturbation expansion breaks down.
But -to repeat - (\ref{two}) was derived
for the regime $p << \alpha $, and the question
now arises whether nothing new happens
from this regime all the way up to $p \rightarrow \infty $, or whether
there is interesting structure at intermediate scales.
In particular, we might envisage a ``quasi-fixed-point'' situation,
in which $g$ remains more or less stationary around the
value $g (0)$ for a wide range of $t$ below $t=0$, before
commencing to grow rapidly at very low momenta.

\subsection{Non-trivial (quasi-)fixed-point
structure at intermediate momenta}
\pr
The answer to the above question turns out
to reside, essentially, in the infrared cutoff $\epsilon$
(which, as we noted above, actually disappeared
from (\ref{two})). The coupling of (\ref{two})
is ``asymptotically free'' (i.e. grows rapidly
in the far infrared)  for $n < 2$, {\it provided}
that the ratio $\alpha / \epsilon $ is large enough
- and in this case dynamical mass generation (d.m.g.)
occurs. To get to the region where d.m.g. does not occur,
we must consider smaller values of $\alpha / \epsilon $,
tending ultimately to unity. This is the region
that will yield the effective
non-trivial fixed point structure.
In this case, $p \simeq \alpha $ and hence
the only allowed
region for the momentum
$k$ in (\ref{sdir}) is $ k \le p$, which now eliminates
the {\it second} term in (\ref{sdir}).
Solving then (\ref{sdir}) in this approximation
(and taking $B = 0$ since d.m.g. does not occur), with
the vertex (\ref{three}), one obtains
\bea
   A(p)&=&1 - \frac{g_0}{3} \int _\epsilon ^ p \frac{dk}{k}
(\frac{k}{p})^3 A^{n-1} (k) = \nn \\
1 -& & \frac{g_0}{3}
      \int _{t_0 - t} ^0             ds e^{3s} A^{n-1} (s)
\label{sdifsol}
\eea
which can be easily solved with the result
\be
   A (t) = ( const + \frac{2-n}{9} g_0 e^{3t_0 -3t})^{\frac{1}{2-n}}
\label{solsd}
\ee
where the $const $ is a positive one and can be found from the
value of the wave function renormalization at $t=
ln(\epsilon/\alpha) \equiv t_0$, namely $A(t_0)
=1$. From
(\ref{sdifsol}) this yields
the value $const  =1 - \frac{2-n}{9}g_0$.
Substituting (\ref{solsd}) back to the gap equation
one obtains a `running' coupling constant in this new
intermediate regime
\be
  g^{I}  \equiv \frac{g_0 e^{3t}}{ (1-\frac{2-n}{9}g_0)
       e^{3t} + \frac{2-n}{9}
g_0 e^{3t_0} }  =
\frac{g_0}{ 1-\frac{2-n}{9}g_0
        + \frac{2-n}{9}
g_0 (\frac{\epsilon}{p})^3 }
\qquad .
\label{renir}
\ee
We note that just as the ``lower scale'' $\epsilon$
disappeared from (\ref{two}), so the ``intermediate scale''
$\alpha $ is absent from (\ref{renir}).
\pr
Let us study the fixed-point structure
of this renormalization-group flow .
To this end, consider the $\beta $ function obtained from
(\ref{renir}) :
\be
 \beta ^I = -dg^I / dt = -3 g^I + \frac{ 3     }{g_0}
(1 - \frac{2-n}{9}g_0)(g^I) ^2
\label{betir}
\ee
Taking into account that $g_0 =\frac{8}{\pi^2 N} $
we observe that the vanishing of $\beta ^I $
occurs not only at $g^I = 0$ but also  at the non-trivial
point
\be
    g^I_* =  \frac{8}{\pi ^2 N} (1 - \frac{2 - n}{9}
\frac{8}{\pi ^2 N} )^{-1}
\label{notriv}
\ee
which indicates the existence of a fixed point
lying at a distance of $O(1/N)$, for $N \rightarrow \infty $,
from the trivial one.
\pr
For what momenta is this fixed point
reached ? Accepting (\ref{renir}) at face value,
the answer would be that it is reached for $p \rightarrow \infty $.
But of course (\ref{renir}) is not valid for $p >> \alpha $,
being appropriate for $\epsilon < p < \alpha $ where
the ratio $\epsilon / \alpha $ is smaller than unity, though not
so very small that $p$ can enter
the region of d.m.g. Referring then to the right hand side
of the second equality in (\ref{renir}), we see that when
$p \simeq \alpha $ the quantity $g^I$ will be very close to $g^I_*$,
differing from it by terms of order $(\epsilon/\alpha )^3\frac{1}{N^2}$,
which is negligible. Indeed, as $p$ moves down to $p \simeq \epsilon$,
$g^I$ arrives at $g_0$, which is still within $(1/N^2)$ of $g^I_*$.
Thus the crucial point is that there is - on the basis of this admittedly
approximate analysis - a significant momentum region
over which the coupling $g^I$
varies very slowly, and we are in a ``quasi-fixed-point'' situation.
In a sense, this slow variation of $g^I$ in the range
$\epsilon < p < \alpha $ (for not too small $\epsilon $)
provides a reconciliation between the normalizations adopted
in the two different approximations (\ref{two}) and (\ref{renir}) -
namely
between $g^L (p=\alpha )= g_0$ and $g^I (p=\epsilon )= g_0$.
\pr
The new fixed point occurs at weak coupling for large $N$.
This is consistent with the interpretation that such a fixed point should
characterize a regime of the theory, as determined by the ratio $\alpha /
\epsilon $, where dynamical mass generation does not occur.
\pr
In summary, then, our analysis suggests a significant
modification of the picture presented
by Kondo and Nakatani~\cite{kondo}. Whereas those authors
only considered $\epsilon << \alpha $, which is the regime
of ``asymptotic freedom'' and d.m.g., we have explored
also the region of smaller values of
$\alpha / \epsilon $, and have concluded that here quantum corrections
create a quasi-fixed-point with weak coupling.
{\it Both} regions of $\alpha / \epsilon $ will be important
in our application of these results to the cuprates,
as we discuss in section 3, where we shall try to relate
the ``$\epsilon $''of this $QED_3$ with the temperature $T$
of $QED_3$ at finite temperature.
\pr
At this stage, it is worth pointing out
the similarity of the above-demonstrated `slow running' of the
effective gauge coupling $g$ at intermediate scales
with (four-dimensional)
particle physics models of `walking technicolour'
type\cite{techni}.
Such models pertain to gauge theories with
asymptotic freedom and involve regions of momentum scale
at which effective running couplings move very slowly
with the scale, exactly as  happens in our (asymptotically
free) $QED_3$ case\footnote{A similarity
of $QED_3$ with walking technicolour
had also been pointed
out previously~\cite{dagotto}, but from a different point of view.
In ref. \cite{dagotto}, a formal analogy
of $QED_3$ with walking technicolour models was noted,
based on the r\^ole
of fermion loops in softening the logarithmic confining gauge potential
to a Coulombic $1/r$ type, in the infrared
regime of momenta. This $1/r$ behaviour of the potential,
and its relevance to dynamical chiral symmetry breaking, is common in both
theories. The formal analogy between $QED_3$ and walking technicolour
theories is achieved~\cite{dagotto}
by replacing the coupling $g^2$ of the four-dimensional
theory
by $1/N$ of $QED_3$.
However $N$ of ref. \cite{dagotto}
does not vary with the energy scale, since wave-function
renormalization effects have not been discussed in their case.
This is the crucial difference in our case, where there is a
more precise analogy
with walking technicolour theories, due to the slowing-down
of the variation of the `effective'
$N$ (\ref{renir}) with the (intermediate) energy scale.}.
This slow running of the coupling
results in such theories in a significant enhancement
of the size of the fermion condensate. In our case,
such condensates are responsible for an opening of a
superconducting gap, and, therefore, one could associate
the slow running of the coupling at intermediate scales
with the suppression of the coherence length of the superconductor
(inverse  of the fermion condensate) in the phase where dynamical
mass generation occurs. Such a suppression, as compared to the
phonon (BCS) type of superconductivity,
which is an experimentally observed and quite
distinctive feature of the
high-$T_c$ cuprates\cite{kirtley},
appears then, in the context of the above gauge theory
model\cite{doreym},
as a natural consequence
of the non-trivial quasi-fixed-point
renormalization group structure. Note that
in ref. \cite{doreym}
the enhancement of the superconducting gap-to-critical-temperature
ratio, as compared to  the standard BCS case, had been attributed
to the super-renormalizability of the theory and the
$T$-independence of quantum corrections, features
which are both associated with the above quasi-fixed-point
(slow running) situation as discussed above.
It is understood, of course, that
before we arrive at definite conclusions about the
actual size of the coherence length in the model,
we should be
able to perform exact calculations by resumming the higher
orders in $1/N$ to see whether these features persist.
At present this is impossible  analytically,
but one could hope for (non-perturbative)
lattice simulations of the above
systems\cite{doreym,kogut}.

\subsection{The Effect of Wave-function Renormalization on the
Effective fermion-fermion interactions and non-trivial (quasi)
fixed points}
\pr
Despite the important physical differences
between the models,
it is worth comparing the above results with
the model of ref. \cite{nayak}, where a
non-trivial
infrared fixed point in the running
of the effective gauge-fermion
coupling was associated with the presence of
a modified
fermion-fermion interaction, of long-range
$1/p^x$ type , with $p$ the momentum.
As mentioned in the Introduction,
the model made explicit use of a P- and T- violating
Chern-Simons interaction for the statistical
gauge field.
The non-relativistic nature of the system of ref.
\cite{nayak} was not important. What was important
was the deviation from the pure Coulombic behaviour
$x =1$, which
itself leads only to marginal deviations from
Fermi-liquid behaviour.
In our case, as we shall argue below,
the r\^ole of $x$ is played
by $1 - O(1/N)$.
The deviation from the Coulombic interactions among fermions
is caused by the non-triviality of $1/N$, and the
(marginal) Coulomb interaction would be recovered in the infinite fermion
flavour limit.
\pr
To make formal
contact with the results of ref. \cite{nayak}
it is essential to compute the (zero-temperature)
effective potential among our fermions, with
wave-function renormalization included.
This is straightforward and we present the result below.
The zero-temperature static potential among fermions
is given by the $\mu=0$ $\nu=0$ form of the gauge-boson
propagator. In ref. \cite{doreym} the effects of the
wave-function renormalization were ignored, which is an accurate
result only in the
$N \rightarrow   \infty$ limit, where the `mean-field' theory
is recovered. This is the Landau fermi-liquid fixed-point.
The $\frac{1}{N}$ corrections yield a non-trivial
wave-function renormalization effect. Resumming the $\frac{1}{N}$
corrections, in an improved  renormalization-group framework,
and
using the ansatz (\ref{three}) for the effective vertex,
we can compute the effective static potential
in a straightforward manner with the result:
\be
      V(p) \propto \frac{\alpha}{8}
 p^{\frac{16n}{3\pi^2 N}}    p^{-1}
\label{nineteen}
\ee
which makes contact with the efective potential
of \cite{nayak} if one identifies
$x = 1 - \frac{16n}{\pi^23N}~<~1$.

\subsection{Comments and comparison with other works}
\pr
Before closing this section
it would be useful to compare our results
with the works
of ref. \cite{appnash,nash}, concerning
existence, as well as gauge invariance,
of a critical number of flavours.
As we have mentioned above, our work does not deal directly
with this issue, which pertains to the infrared momentum regime,
but rather with the
effects of the wave function renormalization at intermediate momenta,
in the presence of an infrared cut-off, which, as we shall argue
below, could be interpreted as expressing
finite temperature effects. In the presence
of an infrared cut-off, a critical number of flavours
has been shown to exist, albeit depending on it~\cite{kondo,aklein}.
The issue of gauge invariance of the result is
still unresolved. The complexity of the
situation can be understood
probably better if we draw an analogy of the
(finite) infrared cut-off with the temperature scale.
In such a case, there are known~\cite{zuk}
unresolved ambiguities
appearing in the low-momentum regime of
the theory, due to non analyticities
of the effective action.
\pr
Below we would like first to compare the results
of ref. \cite{kondo} to those of ref. \cite{appnash,nash}.
In ref. \cite{appnash}, it has been argued, on the basis of
a power-counting analysis, which did not make any use of
the Ward-Takahashi identities, that there is no renormalization
of $N$ to any order in $1/N$,
in the infrared regime of the model.
The arguments were based on the softened Coulombic form of the
gauge-boson propagator in the infrared, as a result of
fermion vacuum polarization : $D_{\mu\nu} \propto
(1/q)(g_{\mu\nu} - (1 - \xi) q_\mu q_\nu /q^2)$, in an arbitrary
$\xi $ gauge, for small momentum transfers $q << \alpha $.
It is worth
noticing, that such arguments
appear to apply equally well to Abelian as well as non-Abelian
theories, since in the latter case
non-Abelian three or four gluon interactions
could not contribute to the potential scaling-violating
interactions.
This analysis has been performed without
implementing an infrared cutoff,
due to the infrared finiteness of the (zero-temperature) theory.
In the work of ref. \cite{kondo}, which is applied
to the infrared regime, an infrared cut-off is
introduced, which changes the scaling properties of the gauge-boson
propagator. In this case,
the scale-invariant situation seems to occur only for
the value $n=2$ in the vertex ansatz, which notably does not
satisfy the Ward-Takahashi identities~\cite{pen}.
As we have seen, gauge invariance requires $n=1$, and in that case
there exists a running $N$, at infrared momentum scales,
as well as
a finite
critical flavour number, which however is
infrared cut-off dependent, and diverges in the limit
where the cut-off is removed.
\pr
We can also compare this result with that of
ref. \cite{nash}, which
claims to have proven the gauge invariance
of the critical number of flavours in $QED_3$.
There, a non-local gauge fixing
was used; this mixes orders in $1/N$ expansion, in the sense that
the gap function in SD contains now graphs of $O(1/N^2)$,
whilst the wave-function renormalization still remains of $O(1/N)$.
In contrast, the analysis
of ref. \cite{kondo} remains consistently
at leading order in $1/N$, and in
the Landau gauge. The meaning of the non-local gauge fixing
is not clear
if one stays consistently within an order
by order $1/N$ expansion. Nor does gauge invariance
make complete sense in the presence of an infrared cutoff.
\pr
Thus, the key to
a possible explanation of the discrepancy between the works of
ref. \cite{appnash,nash} and ref. \cite{kondo} seems to be hidden
in the higher orders in the large $N$ expansion, as
well the presence of the infrared cut-off. Notice that a naive removal
of the infrared cut-off might lead to ambiguities, as becomes
clear from the work of ref. \cite{zuk} for finite temperature
field theories, provided that one makes~\cite{aklein} the
(physically sensible ) identification/analogy
of the infrared cut-off with the temperature scale, at least
within a condensed matter effective theory framework.
\pr
Now we come to
our case.  As we shall argue, our results can offer a way out
of the above-mentioned discrepancy.
For us,
the momentum regime of interest is not the infrared one,
where dynamical mass generation occurs, but the intermediate
scale. In this regime, the power-counting arguments
of ref. \cite{appnash} do not apply, since the gauge-boson
propagator does not have a simple Coulombic behaviour.
Thus, the wave-function renormalization effects, that appear
to exist in our, admitedly rough, truncation of the SD equations,
might not be incompatible with the results of ref. \cite{appnash},
pertaining to the existence of a critical flavour number.
{}From our point
of view, this would mean that,
although there is a (slow) running of an effective $N$,
and thus scale invariance is marginally broken, however,
the running of the coupling
is even more suppressed in the infrared, where strong
quantum effects cut off the increase of the
(asymptotically free) coupling. The infrared cut-off then,
appears as the (spontaneous ?) scale, above which a
slow running of the (asymptotically free) coupling
becomes appreciable. In a condensed-matter-inspired
framework, such a spntaneously appearing scale
makes perfect sense, if one associates the infrared
cut-off with the temperature scale~\cite{aklein}.
For momenta
sightly above the infrared cut-off, then, the situation
of KN ~\cite{kondo} seems to be valid. This regime may be
viewed as the
boundary regime for which dynamical mass generation still
can happen.
Below the infrared scale,
which
is a regime that makes perfect sense in an infrared-finite
theory such as $QED_3$, dynamical mass generation
certainly occurs, and
the arguments of ref. \cite{appnash}
apply, leading to an effective cut-off of the increase
of the coupling constant. In this regime, the gauge-boson
propagator assumes a softened Coulombic $1/r$ form,
which has been argued to be important for a
(superconducting) pairing attraction among fermions (holes)
in the model of ref. \cite{doreym}.
Such a situation, which is depicted
in fig. 1,
was envisaged in
ref. \cite{higashijima} for the case of
chiral symmetry breaking
in four-dimensional $QCD$, which in this way was
de-associated from the confining properties of the theory.
\pr
In the work of KN \cite{kondo} and ours,
all these issues could be confirmed only if
a more complete analysis
of the SD equations, including higher-order $1/N$ corrections,
is performed.
Whether resummation to all orders in $1/N$
washes out completely
the wave-function renormalization effects at intermediate
momenta, leading to an {\it exactly
marginal} (scale invariant) situation,
or keeps this effect at a RG marginal level,
remains an unresolved issue at present.
On the basis of the above discussion,
one would expect that marginal deviations
from scale invariant behaviour at intermediate momenta,
such as the ones
studied in the present work,
survive higher-order analyses,
but they also lead to a critical
number of flavours,
since the latter is an entity pertaining to
the infrared regime of the theory.
Moreover, for us,
who are interested in
performing the analysis in a condensed-matter rather than
particle-theory
framework,
there is
the issue of the ambiguous infrared
limit of the theory at finite temperatures,
which is by no means a trivial matter~\cite{zuk}.
It seems to us that
all these important questions can only be answered
if proper lattice simulations
of the pertinent systems are performed.
At present, the existing computer facilities
might not be sufficient for such an analysis.
\pr
However,
as we shall argue below, the slow running
of the coupling constant of the model
at intermediate momentum scales,
if true,
is a desirable effect
from a condensed matter point of view, where
both infrared and ultraviolet cut-offs should be kept.
The wave-function renormalization effects, discussed above,
prove sufficient in leading to a (marginal) deviation
of the theory from the fermi-liquid fixed point.
At finite temperatures, this effect can have observable
consequences, and might be responsible
for the experimentally observed
abnormal normal state properties
of the high-$T_c$ cuprates, the physics of
which the above gauge theories
are believed to simulate.
We stress once again that such effects
would be absent in an
exactly marginal situation, like the one suggested
in ref. \cite{appnash}.

\section{Linear behaviour of the Resistivity
in $QED_3$ with the temperature scale}
\pr
In this section we want to connect
the above picture of the behaviour
of $QED_3$ at zero temperature
to that of the same theory at finite
temperature, $T$. In the absence, again,
of anything like an exact solution
in the $T \ne 0$ case,
approximations (quite possibly severe ones)
will have to be made.
However, the
physical aim is clear:
we want to connect the experimental observation
that the electrical resistivity
in the normal phase of the high-$T_c$
superconductors varies linearly with $T$
over a wide range in $T$ from low
temperatures up to a scale of $600$ $K$,
to the existence of the non-trivial
quasi-fixed-point structure
of $QED_3$ found in the previous section.
Qualitatively, the way we shall make the
connection is to interpret the temperature
in finite-$T$ $QED_3$ as (related to) an
effective infrared cutoff. This will follow
from the form of the gauge boson propagator
for $T > 0$, to which we now turn.

\subsection{The gauge boson propagator at finite $T > 0$}
\pr
The gauge boson propagator
$\Delta _{\mu\nu} $ is given by the following expression,
\be
  \Delta ^{-1}_{\mu\nu} (p_0, P, \beta ) =
  \Delta ^{(0)-1}_{\mu\nu} (p_0, P, \beta ) +
\Pi _{\mu\nu} (P, p_0, \beta )
\label{fiveb}
\ee
where the vacuum polarization is
given by
\be
 \Pi_{\mu\nu} = \Pi _T (P,p_0, \beta ) P_{\mu\nu} +
\Pi _L (P, p_0, \beta ) Q_{\mu\nu}
\label{fivec}
\ee
The transverse $P_{\mu\nu}$ and longitudinal
$Q_{\mu\nu}$ tensors are given respectively by
\bea
    P_{\mu\nu} = -\delta _{\mu i}(\delta _{ij} - \frac{P_iP_j}{P^2})
\delta _{j\nu} \qquad &;& \qquad Q_{\mu\nu}=-(g_{\mu 0}
- \frac{p_\mu p_0}{p^2})\frac{p^2}{P^2}(g_{0\nu} -
\frac{p_0p_\nu}{p^2}) \nn \\
Q_{\mu\nu} + P_{\mu\nu} &=& g_{\mu\nu} - \frac{p_\mu p_\nu}{p^2}
\label{fived}
\eea
The zero-temperature polarization tensor of the gauge boson is
$\Pi (p,\beta \rightarrow \infty )
=\frac{\alpha p}{8} $. Thus, for low-energies,relevant
for the definition of resistivity , the $p^{-2}$
behaviour of the gauge boson propagator is softened to $p^{-1}$.
For finite temperatures, on the other hand, this behaviour
is softened even more. In the instantaneous
approximation, one finds\cite{inst}
a `longitudinal ' gauge boson mass term
proportional to
\be
       \Pi _{00} (P \rightarrow
       0, p_3 =0,\beta)=\frac{2\alpha ln 2}{\pi \beta }\equiv
2\omega _\pi ^2
\label{eight}
\ee
where $P$ is the magnitude of the spatial
momentum.
Thus we see that, in this approximation,
the temperature has introduced an effective
infrared cutoff $\sim \sqrt{\alpha/\beta}$.
Interpreting this as the ``$\epsilon $''
of the previous section, we find that
the r\^ole of the all-important
ratio $\alpha / \epsilon $
is played by $\sqrt{\beta \alpha}$.
The ``intermediate'' momentum region
is then $\sqrt{\beta\alpha} \gsim 1$, while the
d.m.g. region $\sqrt{\beta \alpha} >> 1$
(or $T << \alpha $).
\pr
In the instantaneous approximation the transverse
gauge bosons remain massless \cite{inst}. However,
beyond the instantaneous approximation~\cite{ait}
one obtains temperature-dependent
corrections also to the transverse parts.
The low-momentum behaviour of these polarization
tensors is not smooth~\cite{ait,zuk}, and in particular
one has the following ambiguities, depending
on the order of the various limits:
\bea
\Pi _L ( P \rightarrow 0, p_3=0, \beta ) &\rightarrow &
2 \omega _\pi ^2                                   \nn \\
\Pi _L ( P =0, p_3 \rightarrow 0, \beta ) &\rightarrow &
 \omega _\pi ^2                                   \nn \\
\Pi _T ( P \rightarrow 0, p_3=0, \beta ) &\rightarrow &
0                  \nn       \\
\Pi _T ( P =0, p_3 \rightarrow 0, \beta ) &\rightarrow &
 \omega _\pi ^2
\label{limits}
\eea
where, in Euclidean formalism,
$p_0$ is replaced by $ip_3$. For our purposes,
however, an approximate form
given in \cite{ait}
will be sufficient:
\be
     \Pi _L \simeq \Pi _T \simeq ( \frac{\alpha p^2}{64}
+ 4\omega _\pi ^4 )^{\frac{1}{2}}
\label{ten}
\ee
where $p^2 = p_3^2 + P^2 $.
In this approximation
the gauge boson propagator reads
\be
\Delta _{\mu\nu} (p) =\frac{g_{\mu\nu} - \frac{p_\mu p_\nu}{p^2}}
{p^2  + \Pi (P,p_3, \beta )}
\label{six}
\ee
where $\Pi $ is given by (\ref{ten}).
In the limit $p \rightarrow 0$, which is relevant
for the definition of resistivity (see below),
one may then replace $\Pi $ by $2 \omega _\pi ^2$,
with the same qualitative association $\epsilon \sim
\sqrt{\alpha/\beta}$ as before. The net
effect of retardation on the gauge boson propagator,
in the large $N$ approximation, is summarized by the form
(\ref{six}).

\subsection{Wave-function Renormalization and Vertex function
at finite $T > 0$ }
\pr
In view of the importance of wavefunction
renormalization in the $T=0$ case, as stressed in Section 2,
it is clear that we must include it also at $T \ne 0$.
We shall find (see below) that its effect is to
provide logarithmic (in $T$) corrections to the linear $T$-dependence
of the resistivity which is characteristic~\cite{lee,ioffe}
of the gauge interactions.
\pr
Wavefunction renormalization
effects in $QED_3$ at $T > 0$ were studied
in \cite{aklein}, using the Pennington-Webb
vertex ansatz ($n=1$ in the notation of (\ref{three})),
and making the
instantaneous approximation,
at least initially.
The approximate SD equation for
$A(P, \beta )$ then becomes (noting
that the ``$A$'' of \cite{aklein} is our $A - 1$)
\be
 A(P, \beta ) \simeq 1 + \frac{\alpha^2}{16N\pi ^2}
\int_0^\alpha dK I(P, K, \beta ) \frac{tanh\frac{\beta}{2}\sqrt{K^2 +
{\cal M}(K, \beta )^2 }}{\sqrt{K^2 + {\cal M}(K, \beta )^2}}
\label{tenb}
\ee
where ${\cal M}$ is the modified
mass function $B/A$, and
\be
I(P,K,\beta )=\frac{K}{\alpha}
\int _0^{2\pi} d\phi \frac{(P^2 - K^2)^2 - Q^4}
{P^2Q^2 [ Q^2 + \Pi_{00} (Q,\beta )^2 ]}
\label{eleven}
\ee
with $Q = |P - K|$.
\pr
However, it was found~\cite{aklein} that the use
of (\ref{tenb}) led to a plainly unphysical
result: {\it viz} $A > 1$. The trouble was traced
to the use of the instantaneous approximation,
which turns out to make a
dramatic impact on $A$, essentially because
of the effective reduction in the dimensionality
of the integration in (\ref{tenb}) from three to two
dimensions.
\pr
An exact treatment is very difficult,
but it was argued in \cite{aklein} that a
plausible improvement to (\ref{tenb}),
taking non-instantaneous terms into account in an approximate
way, is obtained by replacing $\Pi _{00}$
by a $Q$-independent constant $\Delta ^2$ which is
of order $\alpha ^2$, and at the same time
setting the factor $(K/\alpha )$ in (\ref{eleven})
equal to unity.
Certainly the numerical results then obtained, in the region
of dynamical mass gneration, seemed physically sensible :
in particular, as $T \rightarrow 0$, they were in good
qualitative agreement  with previous zero temperature results,
and $A$ was less than unity. In this case, the kernel $I$
is replaced by the temperature-independent
quantity
\be
I_\Delta =-\frac{2\pi}{P^2}\{ 1 -\frac{|P^2 - K^2|}{\Delta ^2}
+ \frac{[P^2 - K^2 + \Delta ^2 ][P^2 - K^2 - \Delta ^2 ]}
{\Delta ^2 \sqrt{[(P-K)^2 + \Delta ^2 ][(P + K)^2 + \Delta ^2 ]} }\}
\label{twelve}
\ee
Although originally discussed, in \cite{aklein},
within a context of dynamical mass generation, the above
approximate formula for $A$ can just as well be
considered in the regime ${\cal M}=0$.
It is for this regime
that we now estimate the resistivity, introducing the
effects of $A$.

\subsection{The resistivity of $QED_3$ in the normal phase}
\pr
Our aim in this subsection
is to exhibit non-fermi liquid behaviour of the resistivity,
and associate it with the quasi-fixed-point structure at
intermediate scales revealed in the previous Section, via the
qualtitative connection $\alpha / \epsilon \sim \sqrt{\beta \alpha}$.
The resistivity of the model is found
by first coupling the system
to an external electromagnetic field $A$
and then computing the response of the effective
action
of the system,
obtained after integrating out the (statistical) gauge
boson and fermion
quanta,
to a change in $A$.
\pr
In the case at hand, in the model of ref. \cite{doreym}
($\tau_3-QED$)
the effective action of the
electromagnetic field, after integrating out hole and
statistical gauge fields\footnote{Due to the $\tau _3$
structure, as a result of the bi-partite lattice structure
\cite{doreym}, there are no cross-terms between the statistical
and the electromagnetic gauge fields to lowest non-trivial
order of a derivative expansion
in the effective action. This implies that in this
model the resistivity is determined by the polarization
tensor of the hole (fermion) loop. On the other hand,
in models where only a single sublattice is used\cite{ioffe,larkin},
such cross terms arise, which are responsible - after the
statistical gauge field integration - for the appearance
of a conductivity tensor
proportional to   $\frac{\Pi _F \Pi _B}{\Pi _B + \Pi _F}$,
with $\Pi _{B,F}$ denoting (respectively) polarization tensors
for the boson fields of the $CP^1$ model and for the fermions (holes)
in a resummed $1/N$ framework.
In such a case, the conductivity is determined by the
lowest conductivity among the subsystems~\cite{larkin}.
In condensed-matter systems
of this type, relevant for the physics of the normal
state of the high-$T_c$ cuprates,
it is the bosonic contribution that determines
the total electrical resistivity~\cite{ioffe}.},
assumes the form
\be
   S_{eff} = \int A^\mu (p) \Delta _{\mu\nu} A^\nu (-p)
\qquad ; \qquad \Delta _{\mu\nu} = (\delta _{\mu\nu} -
\frac{p_\mu  p_\nu}{p^2})\frac{1}{p^2 + \Pi }
\label{conduct}
\ee
in a resummed $1/N$ framework, with $\Pi $ the one-loop
polarization tensor due to fermions.
The functional variation of the effective action
with respect to $A$ yields the electric current $j$.
{}From (\ref{conduct}) this is
proportional to the electric field $E(\omega)= \omega A$,
in, say,  the $A_0 =0$ gauge,
with $\omega$  the energy.
In the normal phase of the electron system,
the proportionality tensor, evaluated at zero spatial momentum,
is $\sigma _f \times \omega$,
with $\sigma _f$
the conductivity\cite{larkin}.
{}From (\ref{conduct}) then,we have
\be
   \sigma _f = \frac{1}{p^2 + \Pi }|_{{\underline P}=0}
\label{polcond}
\ee
where ${\underline P}$
denotes spatial components of the momentum.
\pr
If the effective action were real, then the temperature ($T$)
dependence of the
resistivity
of the model would be given by the $T$-dependence of the
finite-temperature vacuum polarization of the gauge boson.
Thus, following the approximation (\ref{ten}) for the
polarization tensor
in the resummed-$1/N$ framework~\cite{ait},
we would have immediately obtained a
linear $T$-dependence
for the resistivity. Such a temperature dependence
would actually be valid for a wide
range of temperatures above the critical temperature
of dynamical mass generation~\cite{doreym},
due to specific features~\cite{ait} of (\ref{ten}).
\pr
However, things are not so simple.
As first shown by Landau\cite{landau},
the analytic structure of the
vacuum polarization graphs entering the effective action
(\ref{conduct}) is such that there are imaginary parts
in a real-time formalism\cite{weldon}.
These imaginary parts are associated with dissipation
caused by physical processes involving
(on-shell) processes
of the type
{\it fermion } $\rightarrow$ {\it fermion} $+$  {\it gauge
boson}. It turns out that
these constitute the major contributions
to the (microscopic) resistivity\cite{raizer,lee,ioffe}.
In this picture, the latter is
determined by virtue of the Green-Kubo formula\cite{Kubo}
in the theory of linear response, and it turns out to be
inversely proportional to the imaginary part
of the two-point function of the ``electric'' current
$j_\mu^\psi ={\overline \psi } \gamma _\mu \psi $, evaluated at
zero spatial
momentum. In our case,
in the leading $1/N$-resummed framework, the two-point
function of the electric current is given by the graph
of fig. 1. Adopting the ansatz
(\ref{three}) for the vertex function, the result
for the current-current correlator is
\be
  <J_\mu (p) J_\nu (-p) > \propto (A(p))^n \Delta _{\mu\nu} (p)
 (A(p))^n
\label{five}
\ee
To compute
the imaginary parts of (\ref{five})
would require a real-time formalism,
taking into account the processes of Landau damping~\cite{zuk},
which are not an easy matter to compute
in resummed $1/N$ approximation, especially in the limit of
zero-momentum
trasfer, relevant for the definition of
resistivity. Indeed, as shown in ref. \cite{zuk},
and mentioned briefly above,
there is a non-analytic structure
of the imaginary parts of the one-loop polarization tensors
appearing in the quantum corrections of the gauge boson propagator.
Such
non-analyticities result in a non-local effective action.
This non-locality persists
upon coupling the system to an {\it external} electromagnetic
field $ A$. Since the resistivity of the system is
defined
as the response of the system to a variation of $A$,
then the Landau processes, which
constitute the major contribution
to the (microscopic) resistivity,
complicate the situation enormously.
At present, only numerical treatment of these non -analyticities
is possible\cite{zuk,ait}.
\pr
We can circumvent this difficulty, and use only the
real parts of the gauge boson polarization tensor and the
approximate expression (\ref{ten}) to estimate the
temperature dependence of the resistivity,
by
making use of the fact that in ``realistic''
many-body systems~\cite{doreym,lee,ioffe}, believed to be
relevant for a
description of the physics of the cuprates,
there is the phenomenon of spin-charge
separation of the relevant excitations,
discussed briefly in section 4.
According to this picture,
the statistical current (responsible for spin
transport) is opposite to the hole current (electric charge
transport)
\be
 j_\psi + j_z = 0 \qquad; j^\psi _\mu = {\overline \psi }
\gamma _\mu \psi , \qquad j^z_\mu = 2 z^* \partial _\mu z
\label{current}
\ee
and this constraint is implemented by the statistical gauge
field, $a_\mu$, that plays the r\^ole of a Lagrange
multiplier~\cite{ioffe}.
The gauge field, on the other hand, is identified
for physical (on-shell) processes, with the current $j_z$
(of the $CP^1$ model), and thus - on account of (\ref{current}) -
the electric charge is
transported in such systems with a velocity which equals
the propagation velocity $v_F$ of the statistical gauge
fields $a_\mu$\footnote{Again, the model of ref. \cite{doreym}
is different from those of refs. \cite{lee,ioffe}
in that the (independent)
statistical gauge field $a_\mu$ is related (through
its equations of motion) to the sum of the currents
$j_\psi + j_z$. To apply our arguments in this model
one has to assume that for the electric resistivity
the boson part plays no r\^ole, which is
justified by the formula (\ref{polcond}) above. This
allows one to consider only static configurations
for the $z$ fields, thereby justifying the assumption that
the electric charge in the model propagates
with the $a_0$ gauge-boson
velocity.}. In non-trivial vacua, such as the the one
pertaining to our system,
the velocity $v_F$ receives quantum  corrections\cite{pascua}
from vacuum polarization effects.
In a thermal vacuum such corrections are temperature-($T$-) dependent.
\pr
If we represent the (observable) average of the electric current as
$j_\psi = charge \times v_F$, and use Ohm's law to relate it with an
($T$-independent) externally applied electric field
$E$,  $j_\psi = \sigma  . E$, then one observes
that in this picture
the main
$T$-dependence of the resistivity $\sigma ^{-1}$,
comes from
$v_F$, as a result of (thermal) vacuum polarization
effects~\cite{pascua}\footnote{Of course, it is understood
that the above argument is only heuristic and a
proper (microscopic)
computation of the resistivity, using real-time
Green function calculus, combined with kinetic
trasport theory,
appears necessary in order
to arrive at
rigorous results\cite{lee,ioffe}.
However, the heuristic picture above captures
the particular characteristics of the gauge
interactions, responsible for yielding a linear $T$ dependence, as
we show below, and for our purposes it will be sufficient.}.
\pr
To compute $v_F (T)$ we shall use its definition
in the case of
an (on-shell) relativistic massless particle~\cite{pascua}
(in this case the gauge boson)
\be
   v_F = \frac{\partial E}{\partial Q} \qquad ; \qquad
E^2 \equiv q_0^2 = Q^2 + \Pi (Q,\beta)
\label{energy}
\ee
Only the real parts of the gauge boson polarization
tensor are relevant for the computation of (\ref{energy})~\cite{pascua}.
Using (\ref{ten}), it is then straightforward
to evaluate (\ref{energy}) in the limit of vanishing momentum
transfer, appropriate for the definition of resistivity.
The result is
\be
   v_F \propto \frac{Q}{T^{\frac{3}{2}}}  \qquad ; \qquad
   Q \rightarrow \epsilon
\label{veloc}
\ee
Using the association of the momentum infrared
cutoff $Q \simeq \epsilon $ with $\sqrt{\alpha/\beta}
\propto \sqrt{T}$, one gets from (\ref{veloc})
a linear $T$-dependence for $v_F^{-1}$, and thus for the
resistivity $\rho$.
Such a linear $T$ dependence is
a characteristic feature of the gauge
interactions, and, as we shall discuss below,
is valid for a wide range of $T$.
\pr
Above we have ignored
wavefunction renormalization
effects.
We now proceed to include them explicitly,
and demonstrate  the existence of
(logarithmic) deviations
from this linear $T$ behaviour.
This part of the analysis does not require
an explicit computation of the imaginary part
of the correlator (\ref{five}).
It only requires
$A$ evaluated at $p=0$. So we can examine it directly.
In this limit, we have
\be
  I_\Delta (p=0, K) = -\frac{4\pi(\Delta ^2 - K^2)}{(\Delta ^2 + K ^2)^2}
\label{iota}
\ee
The maximum $K$ in (\ref{iota}) runs from $\sim \sqrt{\alpha/\beta}$
to $\sim \alpha $, which in the ``intermediate'' regime
$\alpha / \epsilon \sim \sqrt{\beta\alpha} \gsim 1$
means that $K$ is constrained to lie within
an order
of magnitude of $\alpha$, and that ${\cal M}$ in (\ref{tenb})
will be zero. Recalling that $\Delta ^2$ is also of order
$\alpha ^2$, a rough estimate for $A(p=0, \beta )$
is provided by
\be
A(p=0, \beta) \simeq 1 - \frac{1}{4N\pi}
\int _{\sqrt{\alpha/\beta}}^{\alpha} dK \frac{1}{K} tanh(\beta
\frac{K}{2}) = 1 - \frac{1}{4N\pi} \int _{\sqrt{\alpha\beta}/2}
^{\alpha\beta / 2} \frac{dx}{x} tanh x \qquad .
\label{penultimate}
\ee
\pr
If $\beta \alpha $ were $ >> 1$ (the very low temperature
limit) we could replace the $tanh$ function in (\ref{penultimate})
by unity, and deduce
\be
A(p=0, \beta  ) \simeq 1 - \frac{1}{8N\pi}ln(\alpha\beta) \qquad .
\label{ultimate}
\ee
Then, the resistivity, which formally
is given by the imaginary part
of the inverse of (\ref{five}) as $p \rightarrow 0$, would exhibit
the following temperature dependence (resummed up to $O(1/N)$):
\be
   \rho \propto O(T^{1 - \frac{1}{4N\pi}})
\label{seventeen}
\ee
where we have taken $n=1$ as in \cite{pen}.
We cannot, in any case,
take the precise value of the exponent in (\ref{seventeen})
seriously in view of the rough approximations made along
the way.
\pr
 However the region $\beta \alpha >> 1$
is, in fact, that of dynamical mass generation, rather than
the ``intermediate'' region $\beta \alpha \gsim 1$ in which
we expect the quasi-fixed-point structure to play a r\^ole.
For $\beta \alpha \gsim 1$ the integral of the right hand side of
(\ref{penultimate}) has to be evaluated numerically.
One finds that for $\beta \alpha \gsim 5$ the result
is within $10\%$ of (\ref{ultimate}), and that (\ref{ultimate})
is virtually exact for $\beta \alpha \gsim 10$.
Thus we can conclude that for a wide range of temperature
below $\alpha$, but not so low that the
symmetry-breaking phase is entered, the resistivity
should have the form (\ref{seventeen}), where the precise
coefficient of the $1/N$ power is not known accurately from the
above analysis.
\pr
The main point, then, is the ``stability'' of this $T$-dependence
which correlates remarkably with the quasi-fixed-point structure
of Section 2.

\section{Brief Comments on Realistic models of holons and spinons
for planar doped antiferromagnets}

\subsection{Microscopic models and their (naive) continuum limit}
\pr
Above we have argued that the gauge-fermion interactions
in planar $QED_3$ are responsible for non-ferm-liquid behaviour
in the sense of exhibiting a non-trivial fixed point structure
of the RG at relatively low energies, below the scale
set by the dimensionful coupling constant in three space-time
dimensions.
\pr
The scope of this section is to connect the above results
to realistic models of holons and spinons interacting
magnetically via spin-spin interactions in models
believed to simulate the physics of the recently-discovered
high-$T_c$ materials. We shall be brief and concentrate only
on some heuristic argumentation. Details can be found in the
literature~\cite{bask,shankarspin,doreym2,polch}.
\pr
First we shall identify the r\^oles of the various excitations
of these materials in connection with the various fields
appearing in $QED_3$ models described above.
To this end, we note that in condensed-matter systems, relevant for high-$T_c$
superconductivity, the basic excitations are electron fields
with momenta lying close to the fermi surface.
Optical experiments have shown the existence of a large
fermi surface in these materials. At first sight, this implies that
our model of section 2, based on Dirac fermions, is inadequate.
However, as we remarked earlier, the most important interactions
for fermions, in both the superconducting and the normal phases,
are those which are local on the fermi surface, and as such
an expansion of the effective theory about a single point on this
surface would be adequate. This has been done in ref. \cite{doreym},
with the result that under the {\it assumption} of spin-charge
separation one arrives at an effective low-energy theory
which resembles a variant of $QED_3$, with the Dirac fermions
playing the r\^ole of holon excitations.
\pr
To understand this point, which is our crucial difference from
the approach of refs. \cite{nayak} and \cite{polch} using spinons
only, we remark
that the basic fields are electrons with both spin and charge
described by a creation operator $C_\alpha ^i$, with
$i$ a spatial lattice index, and $\alpha = 1,\dots M$,
a spin $SU(M)$ index. Realistic models have $M=2$.
Spin-charge separation can be implemented by making the
following ansatz~\cite{ioffe,doreym}
\be
             C_\alpha ^i = \psi ^{\dagger,i} z^i_\alpha
\label{spincharge}
\ee
where $\psi^{\dagger,i} $ is a Grassmann field
that represents
the creation of a holon, and $z_\alpha $ is a $CP^{M-1}$ multiplet,
representing a spinon excitation (magnon).
\pr
At this point we
note that in condensed matter physics one uses~\cite{polch,nayak}
an alternative ansatz
\be
            C_\alpha ^i = f^i_\alpha b_i^\dagger
\label{bos}
\ee
where the fermion fields $f$ carry the spin index and
thus represent the spinon excitations,
carrying no electric charge, whilst  the Bose fields
$b^\dagger$ are spinless and are electrically charged.
This is the description followed by \cite{polch},\cite{nayak},
which treats the spin excitations as fermion fields
in the effective lagrangian approach.
This description is related to the
previous one (\ref{spincharge}) by Bosonization tecnhiques and
may be viewed as a
`gauge'-fixing choice~\cite{marchetti}.
\pr
The gauge symmetry in both descriptions can be
found by performing {\it local} phase rotations
of the constituents  in (\ref{spincharge}), (\ref{bos}).
Since for our purposes we shall follow the ansatz (\ref{spincharge})
we concentrate on it from now on.
The Abelian gauge symmetry that leaves the electron field invariant
in (\ref{spincharge}) is
\be
     \psi ^j \rightarrow e^{i\theta(j)} \psi ^j \qquad ; \qquad
z_\alpha ^j \rightarrow e^{i\theta (j)} z_\alpha ^j
\label{abel}
\ee
and is valid beyond half-filling. This gauge symmetry refers to spatial
indices only, and can be expressed in an effective theory formalism
via link variables in a Hartree-Fock approximation~\cite{bask,doreym}
\be
     \sum _{<ij>} \psi ^{i,\dagger} \psi ^j <z_\alpha ^{i,\dagger} z^{\alpha,j}
 >
\equiv \sum _{<ij>} \Delta _{ij} \psi ^{i,\dagger} \psi ^j
\label{gauge}
\ee
where the sums extend over appropriately defined nearest-neighbor
sites to be specified below.
The gauge symmetry is discovered by freezing the
amplitude of the Hartree-Fock field $|\Delta_{ij}| \simeq {\rm const} $,
while letting its phase fluctuate $e^{\int _i^j a.dl} $, with
$a_i$ the spatial components of an Abelian ($U(1)$) gauge field.
\pr
In large-spin approximations~\cite{shankarspin} of doped
antiferromagnets with a bi-partite lattice structure,
intra-sublattice hopping is suppressed by terms of $O(1/S)$, where
$S>>1$ is the effective spin of the excitations.
In this case, the fermion fields in (\ref{spincharge}) $\psi ^i$
may be assigned an internal `colour' quantum number,
labelling the
sublattice they lie on. In such a case the nearest-neighbor
sites in (\ref{gauge}) lie on this sublattice, and from the
point-of-view of the bi-partite lattice are next-to-nearest-neighbors.
The advantage of introducing this bi-partite lattice structure
lies in the fact that the dynamically-generated gap
through the gauge interactions (\ref{gauge}) is parity conserving,
due to energetics in the case of even-flavour fermion
numbers~\cite{nm,doreym,vaf,app}. Thus, one seems to have a natural
explanation of the absence of P,T violation in these materials,
despite the fact that the superconducting (binding) forces are unconventional
(magnetic) in origin.
\pr
The temporal component of the gauge field can be inserted
by invoking
the Gutzwyler projection operator ensuring the
absence of double occupancy in these materials.
This imposes the restriction of {\it at most one electron-per site},
which formally can be expressed via
\be
      \psi ^{i,\dagger}  \psi ^i +  z_\alpha ^{i,\dagger} z_\alpha ^i = 1
\qquad
{\rm no~sum~over~i}
\label{constr}
\ee
In a path-integral approach to quantum
doped antiferromagnets,
the above constraint (\ref{constr})
may be implemented by a Lagrange multiplier
field $a_0$, playing the r\^ole of the temporal
component of the gauge field.
Alternatively, one may work in the $a_0 =0$ axial gauge, appropriately for a
Hamiltonian formulation~\cite{doreym}, in which case one has to use the
constraint explicitly to arrive at an effective lagrangian with the
correct number of independent degrees of freedom.
\pr
In both formulations, the presence of the gauge field indicates
the existence of redundant degrees of freedom which are unphysical.
\pr
The effective lagrangian, describing the
physically-relevant degrees of freedom that lie
close to a single point on the fermi surface can, then, be shown to
acquire the form of a $CP^{1}$ $\sigma$-model, describing the spin
excitations of the system, coupled via a statistical Abelian
gauge field to a system of electrically-charged
Dirac fermions in a spin-charge separated environment,
\be
\frac{1}{\gamma _0} \int d^3x |(\partial _\mu - a_\mu )z|^2
+
\sum _{i=1}^{N}\int d^3x {\overline \Psi}^i (x) (i\nd{\partial}  + \nd{a}\tau_3
- (e/c) \nd{A} )\Psi ^i (x)
\label{efflagr}
\ee
where the constraint
(\ref{constr})
becomes effectively~\cite{doreym2} $z^\dagger z \simeq 1 $.
The quantity
$\gamma _0$ is the antiferromagnetic interaction
coupling constant of the $\sigma$-model~\cite{doreym},
$c$ is the light velocity in units of the fermi velocity
of holes\footnote{For simplicity
we assumed that the fermi velocity of holes is approximately equal to
the velocity of magnons $v_S$ occuring in the $CP^{1}$ sector.
The realistic case is when the two velocities are different,
which spoils the relativistic form of (\ref{efflagr}).
However, this will not be important for our qualitative treatment
in this article. For more comments on this
point see ref. \cite{doreym}.}, $a_\mu$ is the
statistical gauge field, representing magnetic interactions,
and $A_\mu $ is the electromagnetic field. The fermion fields
$\Psi $ are colour doublets, with respect to the sublattice
degree of freedom;
the $\tau _3$ structure,
which acts in this `colour' space,
indicates the opposite spin of the antiferromagnetic
(bi-partite) lattice structure of the underlying lattice.
This doublet structure
should not be confused with the $i=1,2, \dots N$ flavour
degree of freedom of the fields $\Psi$. As we have mentioned
in the introduction, this `flavour number'
represents internal degrees of freedom, associated
with the orientation of the momentum vectors of the
quasiparticle excitations~\cite{gallavoti}
in expansions about a certain
point of a finite-size fermi surface.
For large fermi surfaces, and low-lying (infrared) excitations,
where the cut-off $\Lambda $
effectively collapses to zero, as compared
with the radius $k_F$
of the fermi surface, a controlled large-$N (\Lambda)$ expansion
is then applicable.
\pr
In condensed-matter inspired models~\cite{doreym,qed3T} one
may argue that the spontaneous scale $\alpha$, above which
nothing interesting happens
in $QED_3$~\cite{app}, plays the r\^ole
of the ultraviolet cut-off
$\Lambda$ of ref. \cite{rgshankar}.
Hence, after cell division
of angular space we have
effectively~\cite{rgshankar} $N \sim \alpha /k_F$
(see (\ref{integration}) and following remarks).
In this
interpretation
of the flavour number, which in fact
is essential for a consistent RG approach to
the theory of the fermi surface~\cite{gallavoti},
one has an effective running of the fermion flavour
number with the RG scale, which is precisely the case of our running
$g \propto 1/N$ discussed in section 2.
\pr
To form an
estimate of this effective $N$ we use
the phenomenological formula~\cite{doreym,qed3T,plee}
\be
\alpha = \hbar v_F /(a \eta_{max})
\sim t' (\eta)^{1/2}/\eta _{max}
\label{estimaalpha}
\ee
where $a$ is the lattice spacing,
$v_F$ is the fermi velocity of holes,
$t'$ is a hoping parameter for holes (on the same sublattice),
and $\eta $ $(\eta _{max})$ denotes
the average (maximum
for superconductivity) number density
of holes (doping concentration).
In realistic models the various parameters
entering (\ref{estimaalpha})
depend on temperature, $T$.
For our angular cell division, however, we shall
use the $k_F$ of a zero-temperature theory.
A typical scale for the fermi surface radius, which is
a typical energy of electronic excitations, is thus of
$O [1~eV]$.
For the values of temperature and
doping concentration relevant for
superconductivity
a typical value of $\alpha$ is of order of $eV$~\cite{qed3T}.
As argued in section 3,
in the normal phase, $T > T_c\sim O[100~K]$,
one may replace
the fermi velocity by an effective one
$v_F \propto T^{-1}$, and hence the
corresponding $\alpha (T)$ gets considerably smaller,
as compared to $k_F$,
thereby
shifting the
effective scales towards the infrared, ,
or equivalently pushing the infrared cut-off
to higher values. It is, therefore,
not unreasonable to argue that the conditions for large $N\propto
k_F /\alpha (T) >> 1$ may be satisfied for the
range of temperatures and (large) fermi
momenta characterizing the normal phase
of these materials. Of course, it is understood
that all such estimates are only qualitative.
Any attempt to present quantitatively meaningful considerations
would require working directly with microscopic models, which
falls beyond the scope of the present work.
\pr
Note that for the superconducting phase of the model
the sublattice
structure is important in that the fermion condensate
responsible for the spontaneous breaking of the
electromagnetic gauge invariance $U(1)_{em}$ associated with the
$A$-field in (\ref{efflagr}) occurs between fermions (holes) of opposite
sublattice each of electric  charge $e$.
For the normal phase analysis, however, which we are
interested in for the purposes of the present work,
the sublattice structure is irrelevant.
{}From now on, therefore,
we concentrate on a single sublattice, ignoring
the $\tau_3$ `colour' structure of the fermions.
Whenever the latter becomes important it will be stated explicitly.
\pr
{}From this point of view,
the statistical gauge interaction in (\ref{efflagr}) plays
exactly the r\^ole
of the fermion-gauge interaction
of section 2, that leads to a non-trivial
fixed point structure at momenta $p \lsim O[\alpha] $, where $\alpha$
is the dimensionful scale set by the
statistical gauge interaction coupling constant.
To understand this point it is sufficient to remark that
integrating out the magnon degrees of freedom,
which are massive of mass $m_z$ in the phase
where long-range antiferromagnetic order has been destroyed,
one obtains at low energies (much lower than the mass $m_z$ scale )
a Maxwell-like term for the gauge field $a$ in (\ref{efflagr}), which
thus becomes dynamical~\cite{aitchmav,pol}.
In this sense, the situation for the statistical-gauge
interaction becomes similar to the $QED_3$  case discussed previously.

\subsection{Absence of Charge- or Antiferromagnetic-Density-Wave
Instabilities}
\pr
An interesting question, that arises in connection
with the low-energy behaviour of such systems,
concerns the existence of other type of instabilities
which, from the point of view of an effective lagrangian,
would manifest themselves as marginal or relevant
operators. The obvious class of candidate
interactions, which in fact is the only
one in these models by simple power counting in large-$N$ treatments,
would be four fermion operators.
Since our effective
lagrangian (\ref{efflagr}) has only trilinear
gauge-fermion couplings, such effective operators
could be shown to
arise as a result of ladder (or cross ladder)
graphs involving
the exchange of gauge particles (c.f. fig. 2).
If an operator of this sort is {\it exactly  marginal},
then its scaling would be the same as the tree-level
scaling of the effective gauge-fermion vertex.
Exactly marginal deformations do not cause the appearance
of a gap in the fermion spectrum.
We shall argue below that this is what happens in
our case in the infrared regime of momenta.
\pr
Interesting effects can be examined in this framework
in association with the electromagnetic or statistical
gauge interaction
that could lead to
antiferromagnetic instabilities
in the normal phase, associated with the
formation of electrically-neutral spin (SDW) or charge (CDW)
-density-waves, which
could be described by fermion-antifermion condensates. In our formalism,
since the Grassman variables $\psi^i$
in (\ref{spincharge}) are spinless, the formation of fermion
condensates on a single sublattice would then
be appropriate for a
description of CDW instabilities.
What we shall show below is that in our model such CDW instabilities
cannot occur as a result of the electromagnetic interaction.
Notice that because of the $\tau _3$ structure of our
model (\ref{efflagr}),
the fermion lines in these graphs can all lie
on the same sublattice only if the exchanged gauge particle
is the electromagnetic photon.
Graphs in which the exchanged particle is the statistical gauge boson,
and hence the fermion lines necessarily belong to different
sublatices,
are known~\cite{doreym}
to lead at low momenta to superconducting mass generation
and will not be of interest to us here. In the normal phase
such instabilities are absent.
\pr
Following ref. \cite{polch},
we consider the ladder and cross-ladder graphs
of fig. 2, where the external legs are set to zero momentum, and
the propagators of the electromagnetic (gauge) and fermion fields are dressed
in a Schwinger-Dyson fashion.
The important point for the electromagnetic photon is that
in three dimensions its kinetic term acquires the modified
Coulomb form  (\ref{maxwell}),
in all ranges of momenta; this form implies that the
relevant
propagator scales like
$1/q$, where $q$ is the momentum transfer circulating around the loop
of fig. 2,
for zero external momenta of the fermion legs.
In the phase where there is no gap for the fermion propagators
the latter scales with momenta like
$1/(A(p)\nd{p} )$, where $A(p)$ is the wave-function renormalization.
This is also the same scaling as the one in the region
of momenta $M << p << \alpha $
where dynamical gap generation could occur. Hence for our
purposes we shall adopt this Feynman rule for the
momentum-space scaling of the dressed fermion propagator.
The vertex function is assumed to scale like $A(p)^n\gamma _\mu $
according to the ansatz
(\ref{three}) even for the case of electromagnetic
interactions.
The result of the one loop integral of the ladder and cross-ladder graphs,
then, scales like
\be
    \int d^3q \frac{1}{[q]^4} A^{2n} (q) \times A^{2(n-1)}
\label{ladder}
\ee
Thus, by choosing the Pennington-Webb vertex ansatz, $n =1$,
dictated by gauge-invariance~\cite{pen}, we observe that the
gauge interaction becomes {\it exactly marginal}, since the scaling
behaviour of the ladder and cross-ladder graphs of fig. 2
(\ref{ladder}) is similar to the tree-level scaling, at least in the region
of momenta where dynamical gap generation could occur.
\pr
This implies the absence of charge-density waves
of these systems
caused by the electromagnetic
interactions,
in agreement with more rigorous condensed
matter models~\cite{nonfermi,polch}. It should be remarked that
the above marginal character of the interaction
refers to four-fermion graphs, which from an effective
lagrangian point of view simply denotes the
absence of the pertinent instability caused by such four fermion
interactions. It should not be confused with the
fermion-gauge trilinear
interaction causing a mass gap, which exists anyhow at low
momenta as a result of the gauge interactions~\cite{app,doreym}.
\pr
An additional type
of instability
of such systems is
that of an antiferromagnetic spin-density-wave (SDW).
To study SDW
in the present formalism one should examine
the $CP$-part of the effective action (\ref{efflagr}).
An easier way, which is closer to the present context,
would be to
pass to the alternative spin-charge separation ansatz (\ref{bos}),
by fermionizing the spin excitations. In such a case,
the sublattice structure would be totally irrelevant,
and one should consider the spin degrees of freedom as fermions
interacting with a statistical gauge field of $QED_3$ type.
The low energy behaviour of the system
would be described again
by a modified photon propagator of $1/p$ form,
as a result of fermion vacuum polarisation~\cite{app,pis},
which
would yield exactly marginal four-fermion interactions
as in (\ref{ladder}).
Hence, one finds again that such gauge systems exhibit no
antiferromagnetic instability~\cite{polch}.
\pr
The
masslessness of the gauge particle
was important for the above marginal scaling behaviour,
as was the
modified $1/p$
scaling behaviour of the dressed gauge propagator, which itself
was a result
of the fermion vacuum polarization or (in the case
electromagnetic
interactions) the projection from four to three dimensions~\cite{doreym}.
The fact that the gauge invariance dictates the value $n=1$
in the ansatz (\ref{three}) of the gauge-fermion vertex,
leading to the above marginal behaviour of the
gauge interaction in the ladder graphs of fig. 2,
implies that the absence of charge-density-waves
in the present model,
or antiferromagnetic instabilities in the case of spinon systems,
can be considered as a clear-cut
prediction of the gauge nature of the interactions among the
fermionic quasiparticle excitations.

\subsection{Electromagnetic Effects}
\pr
A final comment concerns the effects of the
electromagnetic field-fermion coupling
on the deviation from fermi-liquid  behaviour in the infrared.
The effect is known to occur
in four space-time dimensions~\cite{vanalphen},
with the result that the
presence of the vector potential in non-relativistic condensed matter
systems  causes deviations from the fermi-liquid
behaviour at low temperatures, which, however, are
suppressed by terms of $O[v_F^2/c^2]$.
\pr
In three space-time dimensions, in the
presence of statistical intractions,
the situation is quite different if
one restricts one's attention in a given
sublattice in these antiferromagnetic oxides.
As
we shall show below,  the
electromagnetic-field-fermion interactions
become irrelevant in the presence of
the electron-electron interactions
caused by the statistical gauge field.
This is easily demonstrated
by first integrating out the auxiliary gauge field
$a_\mu $ in (\ref{efflagr}). We concentrate on the effects of
fermions within each sublattice. In the normal phase, where no mass is
generated, integrating out the fermions
of the other sublattice just produces
Maxwell terms for the statistical gauge field, which due to the
vacuum polarization acquire the form
\be
       {\cal L}^{kin} = \frac{1}{g^2} f_{\mu\nu}^2 + f_{\mu\nu}
\frac{1}{\sqrt{\partial ^2}}f^{\mu\nu} + \dots
\label{sublmax}
\ee
Such terms are irrelevant operators in the infrared, as compared with
the non-derivative $a$-terms in the $CP^{N-1}$  part of the action
(\ref{efflagr}). Indeed,
after $a$-field integration in the sublattice,
one would get current-current terms multiplying the
inverse of the operator
$ {\cal D} _{\mu\nu} = \delta _{\mu\nu} - \frac{\partial ^2 \delta _{\mu\nu} -
\partial _\mu \partial _\nu}{\sqrt{\partial ^2}} $,
appearing in (\ref{sublmax}) .
Only the non-derivative part of such an inverse is relevant in the infrared.
Thus,
reconstructing the electron operators
$\chi $ out of the spin-charge constituents  as~\cite{maruiz}
\be
     \chi _\alpha = z_\alpha ^\dagger  \psi
\label{elec}
\ee
and integrating out the $a$-field in (\ref{efflagr}),
yields a Thirring interaction between the electrically
charged electron fields~\cite{maruiz}
\be
    S^{eff} = \int d^3x [ i {\overline \chi } \nd{\partial} \chi -
\frac{\gamma _0}{4} ({\overline \chi } \gamma _\mu \chi)^2
+ \frac{e}{c} A_\mu {\overline \chi } \gamma ^\mu \chi + \dots ]
\label{electact}
\ee
In the infrared, the electron kinetic terms become irrelevant
operators, as compared with the Thirring contact interactions,
and from now on we shall omit them.
Assuming conservation of the fermion number in each
sublattice\footnote{Spontaneous breaking of the
fermion number occurs in the superconducting
phase, as a result of one-loop anomalies
due to gap generation~\cite{doreym}. In the normal phase, which
we are interested in, such phenomena are absent
and the fermion current is assumed to be conserved at a quantum
level.}, as
a result of the assumed suppression of intrasublattice
and interplanar hopping,
we may represent in three dimensions the
conserved sublattice fermion current as a curl of
a vector field $V_\mu $
\be
    {\overline \chi }\gamma _\mu \chi = \epsilon _{\mu\nu\rho}
\partial _\nu V_\rho
 \label{curl}
\ee
In this case the effective low-energy action (\ref{electact})
can be written in the form
\be
     S^{eff} = \int d^3 x
\frac{e}{c} A_\mu \epsilon ^{\mu\nu\rho}\partial _\nu V_\rho
 -\frac{\gamma _0}{4} F_{\mu\nu} (V)^2 - \frac{\gamma _0}{4} (\partial _\mu
V_\mu )^2  + \dots
\label{vfield}
\ee
where the $\dots $ indicate terms that are more irrelevant, in a RG sense,
in the
infrared, than the terms kept.
The last term in (\ref{vfield}) is viewed as a gauge fixing term.
Our aim is to examine whether the
electromagnetic field interactions
are capable of driving
the theory to a non-trivial fixed point, away from the
free-electron (Landau) fixed point.
We are, thus, interested in the
behaviour of the mixed Chern-Simons term $AdV$ in the presence of
a weak Thirring interaction (i.e. close to the free-electron
(bare) interactions).
This is equivalent to a
strong-coupling problem for the gauge field $V$, which allows a
heuristic proof of the irrelevant character of the $A dV$ interaction, as
follows: first we represent the mixed Chern-Simons  term, in the
infrared, as a heavy-fermion-gauge interaction,
\be
  A.dV \propto  {\overline \Psi } (i\nd{\partial} + \nd{V} \tau _3
+ \frac{e}{c} \nd{A} )\Psi  + M {\overline \Psi}{\Psi }
\qquad M \rightarrow \infty
\label{hevayferm}
\ee
This yields the following two-point
function for the field ${\tilde V} \equiv \epsilon _{\mu\nu\rho}
\partial _\nu V_\rho $:
\be
   K_{\mu\nu} \propto \int d^3x e^{ip.x} <T {\tilde V}(x) {\tilde V}(0) >
= (\delta _{\mu\nu} - \frac{p_\mu p_\nu}{p^2} ) \frac{p^2}
{p^2 + e^2 p^2 I(p) }
\label{2point}
\ee
where
\be
I(p) \propto \frac{1}{4\pi }
(\frac{4M^2}{p^2} )^{\frac{1}{2}}
tan^{-1}[(\frac{4M^2}{p^2} )^{\frac{1}{2}}]
\label{polar}
\ee
with $M \rightarrow \infty $, the auxiliary (massive) fermion mass.
\pr
The scaling of the electromagnetic photon two-point function
is not affected by the $\Psi $ fermions in this limit
and hence it is given by $1/p$, due to (\ref{maxwell}) in three space-time
dimensions.
Thus, we observe that in the infrared
the fermion-current term ${\overline \chi} \gamma _\mu \chi $
is marginal in the sense that it does not scale
with momenta. On the other hand, the
electromagnetic gauge field scales like $p^{-\frac{1}{2}}$,
implying the RG irrelevant nature of the
electromagnetic field-fermion vertex.
\pr
This means that,
in the models examined above, with suppressed
intra-sublattice hopping, in each sublattice
the only dominant deviations from the fermi liquid
behaviour can be induced by the statistical gauge interactions
at energy scales close to $\alpha $.
This result might be subject to experimental
test, provided that accurate enough experiments can be
made so as to obtain data within one sublattice only.
It goes without saying that intra-sublattice hoping, which
increases with increasing doping concentration~\cite{mavsard},
affects the above result.
\pr
\section{Conclusions and Outlook}
\pr
In this article we have examined certain interesting
effects of the wave-function renormalization in (a
variant of)
$QED_3$, which is believed to be a qualitatively correct
continuum limit of semi-realistic condensed
matter systems simulating (planar) high-temperature
superconducting cuprates.
\pr
Based on an (approximate)
Schwinger-Dyson (SD) improved Renormalization Group (RG)
analysis, we have argued for the existence of an (intermediate)
regime of momenta, where the running of the
renormalized dimensionless
coupling of multiflavour
$QED_3$, which is nothing other than the inverse of the
flavour number, is considerably slowed down, exhibiting
a behaviour similar to that of `walking technicolour' models of
particle physics. This slow running, or (quasi) fixed point
structure, has been argued to be responsible
for an increase of the chiral-symmetry breaking
(superconducting) fermion condensate
of the model, as well as for a (marginal) deviation from the
Landau fermi-liquid fixed point. In connection with the latter
property,
we have argued that the large $N$ expansion is fully justified
from a rather rigorous renormalization group approach to
low-energy interacting
fermionic systems with large fermi surfaces.
Some experimentally
observable consequences of this (marginal) non-fermi liquid
behaviour, including
logarithmic temperature-dependent corrections to the
linear resistivity, have been pointed out, which could be
relevant for an explanation of the abnormal normal-state
properties of the high-$T_c$ cuprates.
\pr
The above RG-SD analysis was, however,
only approximately performed at present.
To fully justify the above considerations, and to make
sure that the above-mentioned
effects are not washed out in an exact treatment, one
has to perform lattice simulations of the above models.
Given that this might not be feasible yet, due to the
restricted capacities of the existing computer devices,
an intermediate step would be to perform
a more complete analytic RG treatment
of the relevant large-$N$ SD equations at finite temperatures.
Such a treatment is not easy,
however, due to the mathematical complexity
of the involved equations. In addition, finite-temperature
field theory is known to exhibit unresolved
ambiguities concerning the low momentum limit,
which complicates the situation.
Some of these issues constitute the object of intensive
research effort of our group at present, and we hope to be able
to reach some useful conclusions soon.
\pr
\nk {\Large {\bf Acknowledgements}}
\pr
N.E.M. wishes to thank A. Devoto and the
members of the organizing committee of the
Fourth Chia Meeting on {\it Common Trends in
Condensed Matter and High Energy
Physics}, Chia Laguna, Sardegna (Italy),
3-10 September 1995, for the opportunity they gave him to
present results of this work, and for creating
a very stimulating atmosphere during the meeting.
N.E.M also acknowledges informative discussions with A. Barone,
M.C. Diamantini, G. Semenoff, P. Sodano and C. Trugenberger.

\newpage
\pr
\nk {\Large {\bf Figure captions}}
\pr
\nk {\bf Figure 1}: Running flavour number in $QED_3$.
The coupling is asymptotically free upon the
Pennington-Webb choice
for the vertex function (\ref{three}), corresponding to $n=1$,
as dictated by gauge invariance. The increase of the coupling
is cut-off at the infrared, as a result of the Coulombic form
of the gauge-boson propagator due to fermion vacuum
polarization. Above a certain infrared scale $\epsilon$
the coupling starts running slowly, a situation resembling
that of `walking technicolour'. This kind of behaviour
is argued to be responsible for (marginal) deviations
from the fermi-liquid picture in a condensed-matter
framework.

\pr
\pr
\nk {\bf Figure 2}: Ladder and Cross-Ladder (resummed)
one-loop graphs in $QED_3$.
The soft Coulombic form of the infrared
gauge-boson propagator
results in the exactly marginal character of these
(four-fermion) interactions: the scaling is that
of tree level. This
leads to the absence of the respective
instabilities.

 \end{document}